**Interband electron pairing for superconductivity from the breakdown of the Born-Oppenheimer approximation**

Myung-Hwan Whangbo,*,[a,b,c] Shuiquan Deng,*,[b] Jürgen Köhler[b,d] and Arndt Simon[d]

[a] Department of Chemistry, North Carolina State University, Raleigh, NC 27695-8204, USA
[b] State Laboratory of Structural Chemistry, Fujian Institute of Research on the Structure of Matter (FJIRSM), Chinese Academy of Sciences (CAS), Fuzhou 350002, China
[c] State Key Loboratory of Crystal Materials, Shandong University, Jinan 250100, China
[d] Max-Planck-Institute for Solid State Research, Heisenbergstr. 1, D-70569, Stuttgart, Germany

**Abstract**

The origin of interband electron pairing responsible for enhancing superconductivity and the factors controlling its strength were examined. We show that the interband electron pairing is a natural consequence of breaking down the Born-Oppenheimer approximation during the electron-phonon interactions. Its strength is determined by the pair-state excitations around the Fermi surfaces that take place to form a superconducting state. Fermi surfaces favorable for the pairing were found and its implications were discussed.

Since the discovery of superconductivity in Hg at 4 K in 1911,[1] numerous studies have been carried out to find other superconductors with higher superconducting transition temperature $T_c$ and understand the cause for superconductivity. The charge carriers of superconductors are pairs of electrons while those of normal metals are individual electrons. The Bardeen-Cooper-Schrieffer (BCS) theory of superconductivity,[2] in which the electron pairing arises from electron-phonon interactions, showed that the $T_c$ increases with raising the average phonon frequency $\langle\omega\rangle$ of the lattice, the electronic density of states (DOS) at the Fermi level $n(e_F)$ and the electron-phonon coupling constant $\lambda$. By explicitly taking into consideration the actual electron-phonon interactions and the Coulomb repulsion between pairing electrons, Eliashberg extended the BCS theory to show the relationship of $T_c$ to the effective Coulomb repulsion $\mu^*$ and the electron-phonon spectral function $\alpha^2(\omega)F(\omega)$.[3,4] McMillan numerically solved the Eliashberg equations as a function of $\mu^*$ and $\alpha^2(\omega)F(\omega)$ in a small range of $\mu^*$ and $\lambda$ to express $T_c$ as an exponential function of $\lambda$ and $\mu^*$ with the prefactor $\Theta_D/1.45$, where $\Theta_D$ is the Debye temperature.[5] The McMillan equation was improved by Allen and Dynes[6] by modifying the prefactor such that all parameters constituting the prefactor can be calculated once the spectral function $\alpha^2(\omega)F(\omega)$ is known. Thus, with typical value of $\mu^* = 0.13$, one can predict the $T_c$ of any metal on the basis of the Allen-Dynes modified McMillan equation by evaluating the spectral function $\alpha^2(\omega)F(\omega)$ in terms of first principles electronic structure calculations. This quantitative approach led Duan et al.[7] in 2014 to predict that $H_2S$ under 200 GPa would become a superconductor at $T_c = 191 - 204$ K. The prediction was quickly verified by Drozdov et al. in 2015,[8] who showed that hydrogen sulfide, $H_2S$, becomes a superconductor at $T_c$ = 203 K under a pressure of 153 GPa.

It had been believed that the McMillan equation has an upper limit of $T_c$ even if $\lambda \to \infty$, and this McMillan limit was thought to be 28 K until the discovery of high-$T_c$ cuprate superconductors.[9] However, the McMillan equation was obtained by numerically solving the Eliashberg equation in a small range of $\lambda$ (< 2). The Eliashberg equation for $\lambda \gg 1$ reveals[4] that $T_c \propto \sqrt{\lambda}$, so this theory places no upper bound of $T_c$ for conventional superconductors, i.e., those based on the electron-phonon mechanism of superconductivity. However, in the strong coupling limit ($\lambda \gg 1$), a metallic system may enter a polaronic regime.[10] The first class of superconductors breaking the McMillan limit is the layered cuprates; $La_{2-x}Ba_xCuO_4$ with $T_c$ = 35 K found in 1986,[9] $YBa_2Cu_3O_{7-x}$ with $T_c$ = 93 K in 1987,[11] and $HgBa_2Ca_2Cu_3O_{8+x}$ with $T_c$ = 134 K in 1993.[12] The $T_c$ of $HgBa_2Ca_2Cu_3O_{8+x}$ is raised to 153 K under pressure of 150 kbar.[13] Another class of superconductors breaking the McMillan limit is the layered ferropnictides, with the highest $T_c$ = 55 K found for $SmFeAsO_{1-x}F_x$ in 2008.[14] It has been generally believed that the cuprate and ferropnictide superconductors appear unconventional, namely, their electron-pairing mechanisms differ from electron-phonon coupling. However, it has recently been found[15] that the high-$T_c$ cuprate $YBa_2Cu_3O_{7-\delta}$ adopts vortex states in a magnetic field, a signature predicted for conventional type-II superconductors by the BCS theory, suggesting that the superconductivity of cuprate superconductors may turn out to be conventional.[15,16] The layered compound $MgB_2$, found to be superconducting at 39 K in 2001,[17] is a conventional superconductor breaking the McMillan limit. To explain the superconducting properties of $MgB_2$, it is necessary to consider that $MgB_2$ has two different superconducting energy gaps, namely, with two partially-filled bands.[18-20] As already mentioned, the highest $T_c$ of 203 K has been discovered for $H_2S$ albeit under extremely high pressure.[8]

Within the BCS theory, a superconductor with $T_c$ well above the McMillan limit is explained as due to a "multiband" mechanism, supposing the presence of more than one partially-filled band.[21-24] In such a case, the electron pairing can occur not only within each band (i.e., the intraband electron pairing) but also between different bands (i.e., the interband electron pairing). Studies based on a model Hamiltonian have shown that $T_c$ of a conventional superconductor can be raised significantly if the Hamiltonian includes the interband electron pairing term.[24] So far, however, no study has shown why the interband electron pairing occurs and how it controls the strength of interband pairing. In the present work, we explore these questions by studying the breakdown of the Born-Oppenheimer approximation (BOA),[25,26] which takes place during the electron-phonon interaction, and the electron-pair excitations around the Fermi surfaces leading to a superconducting state. We show that the interband electron pairing results from the BOA breakdown, and the shapes of the Fermi surfaces control the strength of the interband electron pairing.

A metal can lower its energy by introducing a bandgap at the Fermi level in two typical ways; one is to form a charge density wave (CDW) state[27,28] and the other is to form a superconducting state.[29] In both cases, the energy lowering involves the interaction of the occupied states with the unoccupied states lying close to the Fermi level, but they differ in the way these states interact.[28,29] The formation of a CDW induces a metal-to-insulator phase transition, but that of a superconducting state a metal-to-superconductor phase transition. To illustrate the difference between the CDW and superconducting states, we consider for simplicity a partially-filled one-dimensional (1D) band, with the dispersion relation, e(**k**) vs. **k**, shown in **Figure 1a**. In general, e(**k**$\sigma$) = e(-**k**$\sigma$), where $\sigma$ refers to the spin (i.e., $\sigma$ = ↑ or ↓). Without loss of generality, one can assume that the states $\phi$(**k**$\sigma$) are occupied in the k-space region of -**k**$_F$ $\leq$ **k** $\leq$ **k**$_F$, where **k**$_F$ is the Fermi vector for which e(**k**$_F\sigma$) = e$_F$, while those outside this region are unoccupied. For convenience, the wave vectors leading to the occupied and unoccupied states may be referred to as **k**$_o$ and **k**$_u$, respectively. Then, e(**k**$_o\sigma$) $\leq$ e$_F$, and e(**k**$_u\sigma$) $\geq$ e$_F$. At **k**$_F$, the occupied and unoccupied states are degenerate. The Fermi surface, being the boundary surface in k-space between the **k**$_o$ and **k**$_u$ regions, is given by two points, -**k**$_F$ and **k**$_F$ in the 1D representation of k-space (**Figure 1a**), and by the two parallel lines perpendicular to the Γ−X line in a two-dimensional (2D) representation of k-space (**Figure 1b,c**), and by the two parallel planes perpendicular to the Γ−X line in a three-dimensional (3D) representation of k-space (not shown).

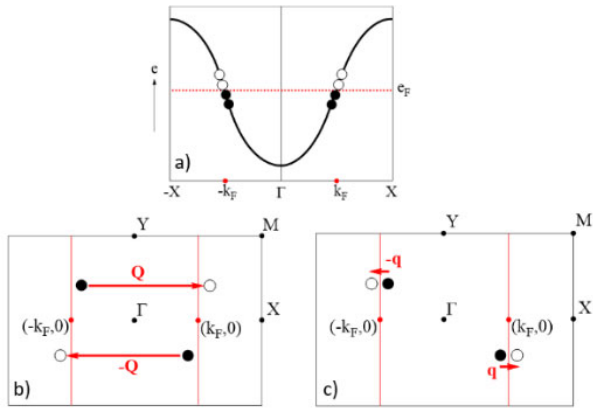

Figure 1. a) 1D representation for the dispersion relation of a partially-filled 1D band, where the filled and empty circles represent the occupied and unoccupied states around the Fermi level, respectively. b,c) 2D representation of the Fermi surface (red lines) of the partially-filled 1D band. One-electron excitations leading to a CDW state are illustrated in b), and electron-pair excitations leading to a superconducting state in c).

The states of a metal important for energy lowering via the bandgap opening at the Fermi level are those lying close to the Fermi level, that is, their wave vectors **k**$_u$ and **k**$_o$ are close to $\pm$**k**$_F$ (**Figure 1b**). A CDW state with propagation vector **Q** = 2**k**$_F$ is formed by the interaction of the occupied states $\phi$(**k**$_o\sigma$) with the unoccupied states $\phi$(**k**$_u\sigma$), where **k**$_u$ and **k**$_o$ are chosen such that **k**$_u$ - **k**$_o$ = $\pm$**Q** (**Figure 1b**) with **Q** representing the lattice phonon. In the second quantization terminology, the interaction of $\phi$(**k**$_o\sigma$) with $\phi$(**k**$_u\sigma$) is described as the excitation from $\phi$(**k**$_o\sigma$) to $\phi$(**k**$_u\sigma$), namely, by the term $c^+$(**k**$_u\sigma$)$c$(**k**$_o\sigma$), where $c^+$(**k**$_u\sigma$) and $c$(**k**$_o\sigma$) are the creation and annihilation operators of an electron in the state $\phi$(**k**$_u\sigma$) and $\phi$(**k**$_o\sigma$), respectively. The tendency toward a CDW formation becomes strong when the Fermi surface meets the nesting condition as shown in **Figure 1b** (by the vector $\pm$**Q**).[27-29] (For further discussion of a CDW state, see ref. 28c.)

For conventional superconductors, the electron pairing is mediated by electron-phonon coupling. In a crystalline lattice made up of cations, an electron moving in a certain direction momentarily drags the nearby cations (**Figure 2a**). Then, another electron moving in the opposite direction perceives the displacements of the cations and is attracted to them. Consequently, the two electrons move effectively as a pair as if they are coupled by an attractive force. The energy lowering that leads to a superconducting state arises from the interaction of occupied pair-states $\phi(+\mathbf{k}_o\uparrow)\phi(-\mathbf{k}_o\downarrow)$ with unoccupied pair-states $\phi(+\mathbf{k}_u\uparrow)\phi(-\mathbf{k}_u\downarrow)$,[29] where $\mathbf{k}_o$ and $\mathbf{k}_u$ lie close to $\pm\mathbf{k}_F$(**Figure 1c**). The conservation of the total momentum of the electron pair requires that $\mathbf{k}_u - \mathbf{k}_o = \pm\mathbf{q}$ for one electron (**Figure 2b**), where $\mathbf{q}$ represents the lattice phonon. In the second quantization terminology, the interaction of the occupied pair-state $\phi(+\mathbf{k}_o\uparrow)\phi(-\mathbf{k}_o\downarrow)$ with the unoccupied pair-state $\phi(+\mathbf{k}_u\uparrow)\phi(-\mathbf{k}_u\downarrow)$ is described as the excitation from $\phi(+\mathbf{k}_o\uparrow)\phi(-\mathbf{k}_o\downarrow)$ to $\phi(+\mathbf{k}_u\uparrow)\phi(-\mathbf{k}_u\downarrow)$, i.e., by the term $c^+(+\mathbf{k}_u\uparrow)c^+(-\mathbf{k}_u\downarrow)c(-\mathbf{k}_o\downarrow)c(+\mathbf{k}_o\uparrow)$. For the convenience of our discussion, we introduce the simplified notations for the pair-states,

$$\phi(+\mathbf{k}_o\uparrow)\phi(-\mathbf{k}_o\downarrow) \Rightarrow (+\mathbf{k}_o\uparrow)(-\mathbf{k}_o\downarrow),$$
$$\phi(+\mathbf{k}_u\uparrow)\phi(-\mathbf{k}_u\downarrow) \Rightarrow (+\mathbf{k}_u\uparrow)(-\mathbf{k}_u\downarrow).$$

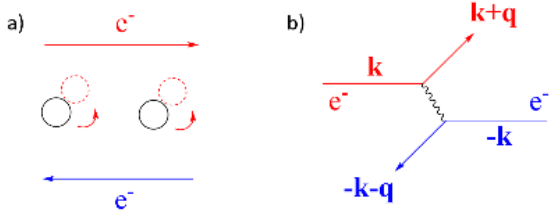

Figure 2. Illustration of the electron-phonon interactions leading to electron pairing in a) real and b) momentum space. The wiggly line in b) represents the electron-phonon interaction.

The interaction between $(+\mathbf{k}_o\uparrow)(-\mathbf{k}_o\downarrow)$ and $(+\mathbf{k}_u\uparrow)(-\mathbf{k}_u\downarrow)$ results in mixed pair-states,

$$\alpha(\mathbf{k})(+\mathbf{k}_o\uparrow)(-\mathbf{k}_o\downarrow) + \beta(\mathbf{k})(+\mathbf{k}_u\uparrow)(-\mathbf{k}_u\downarrow),$$

where $\alpha(\mathbf{k})$ and $\beta(\mathbf{k})$ are the mixing coefficients. Such states are generated for various combinations of $\mathbf{k}_o$ and $\mathbf{k}_u$ states. Then, the superconducting state is described by the product of such mixed pair-states,

$$\Pi_\mathbf{k}[\alpha(\mathbf{k})(+\mathbf{k}_o\uparrow)(-\mathbf{k}_o\downarrow) + \beta(\mathbf{k})(+\mathbf{k}_u\uparrow)(-\mathbf{k}_u\downarrow)].$$

A similar notation was also used by De Gennes to describe the superconducting state.[29b] In this notation, $(+\mathbf{k}_o\uparrow)(-\mathbf{k}_o\downarrow)$ and $(+\mathbf{k}_u\uparrow)(-\mathbf{k}_u\downarrow)$ represent the occupied and unoccupied cases of a paired state $(+\mathbf{k}\uparrow)(-\mathbf{k}\downarrow)$, respectively. In the present work, we are interested only in the pair-state excitations from $\phi(+\mathbf{k}_o\uparrow)\phi(-\mathbf{k}_o\downarrow)$ to $\phi(+\mathbf{k}_u\uparrow)\phi(-\mathbf{k}_u\downarrow)$ in the vicinity of the Fermi level from the perspective of Fermi surfaces.

For simplicity, we consider a metal with two partially-filled bands. In general, the electronic band structure of a metal is determined under the BOA.[25,26] We represent bands 1 and 2 obtained under this approximation as $\phi_1^0(\mathbf{R}, \mathbf{k}\sigma)$ and $\phi_2^0(\mathbf{R}, \mathbf{k}\sigma)$, respectively, where $\mathbf{R}$ represents the nuclear framework.

During the process of electron-phonon coupling (**Figure 2**), the BOA is no longer valid because the electron movement induces, though momentarily, a displacement of the ions from their equilibrium positions. This BOA breakdown mixes $\phi_1^0(\mathbf{R}, \mathbf{k}\sigma)$ and $\phi_2^0(\mathbf{R}, \mathbf{k}\sigma)$ to form new states, $\psi(\mathbf{R}, \mathbf{k}\sigma)$, which may be written as[26]

$$\psi(\mathbf{R}, \mathbf{k}\sigma) = \Omega(\mathbf{R}, \mathbf{k})[C_1\phi_1^0(\mathbf{R}, \mathbf{k}\sigma) + C_2\phi_2^0(\mathbf{R}, \mathbf{k}\sigma)] \qquad (1)$$

where $\Omega(\mathbf{R}, \mathbf{k})$ represents the vibrational wave function. The BOA breakdown[25,26] leads to the energy lowering that involves the gradient of the electronic wave functions $\phi_i^0(\mathbf{R}, \mathbf{k}\sigma)$ (i = 1, 2) as well as that of the nuclear wave function $\Omega(\mathbf{R}, \mathbf{k})$. This energy lowering, $\Delta E_{BD}$, associated with the BOA breakdown, involves changes in both the electronic and the nuclear wave functions,[26]

$$\Delta E_{BD} = -\sum_I \frac{1}{M_I} \int \Omega(\mathbf{R}, \mathbf{k})\nabla_I\Omega(\mathbf{R}, \mathbf{k}) \cdot \langle\phi_2^0(\mathbf{R}, \mathbf{k}\sigma)|\nabla_I|\phi_1^0(\mathbf{R}, \mathbf{k}\sigma)\rangle d\mathbf{R} \qquad (2)$$

where $M_I$ and $\nabla_I$ are the mass and the gradient of the nucleus with mass M, respectively. (The momentum of a nucleus M is given by $-i\nabla_I$ in atomic unit, so Eq. 2 shows that the interaction between the electronic states changes the momenta of the cations, which in turn induces the cation displacements.) The $\langle\phi_2^0(\mathbf{R}, \mathbf{k}\sigma)|\nabla_I|\phi_1^0(\mathbf{R}, \mathbf{k}\sigma)\rangle$ term can be large when $\phi_2^0(\mathbf{R}, \mathbf{k}\sigma)$ and $\phi_1^0(\mathbf{R}, \mathbf{k}\sigma)$ are degenerate.[26] (In the absence of explicit expressions for $\phi_2^0(\mathbf{R}, \mathbf{k}\sigma)$ and $\phi_1^0(\mathbf{R}, \mathbf{k}\sigma)$, there is no symmetry argument with which to find whether the matrix element $\langle\phi_2^0(\mathbf{R}, \mathbf{k}\sigma)|\nabla_I|\phi_1^0(\mathbf{R}, \mathbf{k}\sigma)\rangle$ vanishes or not. Here we assume that, for most cases of $\mathbf{k}$, the matrix elements $\langle\phi_2^0(\mathbf{R}, \mathbf{k}\sigma)|\nabla_I|\phi_1^0(\mathbf{R}, \mathbf{k}\sigma)\rangle$ do not vanish even though they may be small in magnitude. Otherwise, the interactions between $\phi_2^0(\mathbf{R}, \mathbf{k}\sigma)$ and $\phi_1^0(\mathbf{R}, \mathbf{k}\sigma)$ lead to no energy lowering.) This condition is met around the Fermi level so that the BOA breakdown is most strongly felt by the states lying around the Fermi level.

To consider the pair-states arising from $\psi(R, k\sigma)$, we introduce the simplified notation:

$$\psi(\mathbf{R}, +\mathbf{k}\uparrow)\psi(\mathbf{R}, -\mathbf{k}\downarrow) \Rightarrow (+\mathbf{k}\uparrow)(-\mathbf{k}\downarrow).$$

Similarly, the pair-states arising from the $\phi_i^0(\mathbf{R}, \mathbf{k}\sigma)$ (i = 1, 2) can be simplified as:

$$\phi_i^0(\mathbf{R}, +\mathbf{k}\uparrow)\phi_i^0(\mathbf{R}, -\mathbf{k}\downarrow) \Rightarrow (i^0, +\mathbf{k}\uparrow)(i^0, -\mathbf{k}\downarrow)$$

Then, according to Eq. 1, the occupied pair-states associated with $\psi(\mathbf{R}, \mathbf{k}\sigma)$ are written as

$$\begin{aligned}&(+\mathbf{k}_o\uparrow)(-\mathbf{k}_o\downarrow)\\&= [C_1(1^0, +\mathbf{k}_o\uparrow) + C_2(2^0, +\mathbf{k}_o\uparrow)][C_1(1^0, -\mathbf{k}_o\downarrow) + C_2(2^0, -\mathbf{k}_o\downarrow)]\\&\approx C_1^2(1^0, +\mathbf{k}_o\uparrow)(1^0, -\mathbf{k}_o\downarrow) + C_2^2(2^0, +\mathbf{k}_o\uparrow)(2^0, -\mathbf{k}_o\downarrow)\end{aligned} \qquad (3)$$

In obtaining the second equality, the occupied or unoccupied pair-states formed between two different bands $\phi_1^0(\mathbf{R},\mathbf{k}\sigma)$ and $\phi_2^0(\mathbf{R},\mathbf{k}\sigma)$ are neglected. Similarly, the unoccupied pair-states associated with $\psi(\mathbf{R},\mathbf{k}\sigma)$ are written as

$$(+\mathbf{k}_u\uparrow)(-\mathbf{k}_u\downarrow)$$
$$= [C_1(1^0,+\mathbf{k}_u\uparrow) + C_2(2^0,+\mathbf{k}_u\uparrow)][C_1(1^0,-\mathbf{k}_u\downarrow) + C_2(2^0,-\mathbf{k}_u\downarrow)]$$
$$\approx C_1^2(1^0,+\mathbf{k}_u\uparrow)(1^0,-\mathbf{k}_u\downarrow) + C_2^2(2^0,+\mathbf{k}_u\uparrow)(2^0,-\mathbf{k}_u\downarrow) \qquad (4)$$

Then, the pair-state excitation from $(+\mathbf{k}_o\uparrow)(-\mathbf{k}_o\downarrow)$ to $(+\mathbf{k}_u\uparrow)(-\mathbf{k}_u\downarrow)$ induces not only the intraband excitations

$$(1^0,+\mathbf{k}_o\uparrow)(1^0,-\mathbf{k}_o\downarrow) \rightarrow (1^0,+\mathbf{k}_u\uparrow)(1^0,-\mathbf{k}_u\downarrow)$$
$$(2^0,+\mathbf{k}_o\uparrow)(2^0,-\mathbf{k}_o\downarrow) \rightarrow (2^0,+\mathbf{k}_u\uparrow)(2^0,-\mathbf{k}_u\downarrow)$$

but also the interband excitations

$$(1^0,+\mathbf{k}_o\uparrow)(1^0,-\mathbf{k}_o\downarrow) \rightarrow (2^0,+\mathbf{k}_u\uparrow)(2^0,-\mathbf{k}_u\downarrow)$$
$$(2^0,+\mathbf{k}_o\uparrow)(2^0,-\mathbf{k}_o\downarrow) \rightarrow (1^0,+\mathbf{k}_u\uparrow)(1^0,-\mathbf{k}_u\downarrow)$$

In the intraband electron pairing, all the occupied and unoccupied wave vectors involved in the pair-state excitations lie close to the Fermi surface, with the wave vector **q** created from the momentary cation displacements of the lattice. For the pair-state excitations between two different Fermi surfaces, the situation becomes more complicated. For the two Fermi surfaces that cross each other as shown in **Figure 3a**, for example, the magnitudes of the differences $\mathbf{k}_u$ - $\mathbf{k}_o$ for the interband excitations can be small only for those $\mathbf{k}_u$ and $\mathbf{k}_o$ chosen from the immediate vicinity of the crossing points but vary widely depending on the regions of the k-space. In such a case, the varying magnitudes of $\mathbf{k}_u$ - $\mathbf{k}_o$ can only be supplied by specific momentary cation displacements of the lattice so that such an interband electron pairing will be less effective. There are two cases when the interband electron pairing can become significant as discussed below.

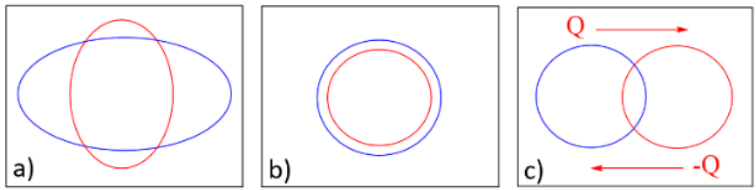

Figure 3. a) Two different 2D Fermi surfaces leading to an ineffective interband electron pairing. b,c) Two different 2D Fermi surfaces leading to effective interband electron pairing. In each case, different bands are represented with different colors. The two Fermi surfaces cross each other in a limited number of k-points in a), do not overlap but are similar in shape and are close to each other in b), and are nested in c).

One occurs when two Fermi surfaces are not identical but are very similar as depicted in **Figure 3b** so that the wave vector difference $\mathbf{k}_u$ - $\mathbf{k}_o$ between the two Fermi surfaces becomes small enough to be supplied by the momentary displacements of the cations. The latter can become substantial if the lattice has a large average phonon frequency $\langle\omega\rangle$, which occurs when the phonons involve the vibrations of light atoms. The best candidate for such a case is found for $MgB_2$, which has two nearly-overlapping cylindrical Fermi surfaces oriented along the $\Gamma \rightarrow A$ direction.[23,30]

The other case occurs when the two Fermi surfaces are nested by a vector **Q** as depicted in **Figure 3c**, so that all the wave vector differences $\mathbf{k}_u$ - $\mathbf{k}_o$ associated with the interband pair-state excitations are given by $\pm\mathbf{Q}$ = $\mathbf{k}_u$ - $\mathbf{k}_o$. A metallic system possessing a nested Fermi surface tends to undergo a CDW formation, leading to the periodic lattice distortion associated with the nesting vector **Q**.[27-29] A metal is expected to survive the CDW instability if the lattice does not provide phonons to couple with the electronic structure to open a bandgap at the Fermi level. In such case, the nested Fermi surface can be used for interband pair-state excitations. This situation applies to the high $T_c$ superconductivity of iron-based superconductors, which were found to possess nested Fermi surfaces.[31-34]

Note that, in a 3D representation of k-space, each Fermi surface of **Figure 3**, represented by a circle or an ellipse, becomes a cylinder running along the $\Gamma \rightarrow Z$ direction. In general, the Fermi surfaces of a 3D metal are rather complicated so that, when two such surfaces interpenetrate, there are not many k-points belonging to both surfaces. Thus, in terms of generating effective interband pair-state excitations, metals of 2D layered structures provide more favorable Fermi surfaces than do those of 3D structures. This might explain in part why high $T_c$ superconductors are mostly found in metals of layered structures. In summary, the interband electron pairing results naturally from the BOA breakdown. The Fermi surfaces favorable for interband pair-state excitations are more likely found for 2D layered metals than for 3D metals. The interband pair-state excitations become favorable when two Fermi surfaces are very similar or nested and when the metal has a large average phonon frequency $\langle\omega\rangle$.

**Acknowledgements**

MHW thanks Annette Bussmann-Holder for discussions on multiband superconductivity. This work was supported by National Natural Science Foundation of China (61874122, 21703251), National Key Research and Development Program of China (2016YFB0701001), The Strategic Priority Research Program of the Chinese Academy of Sciences (XDB20000000) and 100 talents program of CAS and Fujian Province.

**Keywords:** Interband electron pairing • Multiband superconductivity • Breakdown of the Born-Oppenheimer approximation